# Modified Newton-Raphson GRAPE methods
# for optimal control of spin systems

D.L. Goodwin, Ilya Kuprov[*]

*School of Chemistry, University of Southampton,*
*Highfield Campus, Southampton, SO17 1BJ, UK.*

[*]Corresponding author (i.kuprov@soton.ac.uk)

**Abstract**

Quadratic convergence throughout the active space is achieved for the gradient ascent pulse engineering (GRAPE) family of quantum optimal control algorithms. We demonstrate in this communication that the Hessian of the GRAPE fidelity functional is unusually cheap, having the same asymptotic complexity scaling as the functional itself. This leads to the possibility of using very efficient numerical optimization techniques. In particular, the Newton-Raphson method with a rational function optimization (RFO) regularized Hessian is shown in this work to require fewer system trajectory evaluations than any other algorithm in the GRAPE family. This communication describes algebraic and numerical implementation aspects (matrix exponential recycling, Hessian regularization, *etc.*) for the RFO Newton-Raphson version of GRAPE and reports benchmarks for common spin state control problems in magnetic resonance spectroscopy.

**Introduction**

Scientific instruments used in many applications of quantum theory have reached the limits of what is physically, legally or financially possible. Examples include power deposition safeguards in MRI instruments [1], sample heating thresholds in biomolecular NMR spectroscopy [2] and the steep dependence of the cost of superconducting magnets on the induction they generate [3]. Some limits, such as the length of time a patient can be persuaded to stay inside an MRI machine, are psychological [4], but in practice no less real.

The art of running an experiment to a given accuracy with minimal expenditure of time and resources is known as *optimal control theory* [5,6]. In quantum mechanics a popular formulation [7,8] is to specify the desired state vector $\boldsymbol{\delta}$ of the system and maximise the following fidelity functional $J$ with respect to the instrumentally controllable part of the Hamiltonian $\mathbf{H}_1(t)$:

$$J\left[\mathbf{H}_1(t)\right] = \mathrm{Re}\langle\boldsymbol{\delta}|\exp_{(O)}\left[-i\int_0^T \left(\mathbf{H}_0 + \mathbf{H}_1(t) + i\mathbf{R}\right)dt\right]|\boldsymbol{\rho}_0\rangle - J_{\mathrm{RF}}\left[\mathbf{H}_1(t)\right] \qquad (1)$$

where $\exp_{(O)}$ indicates a time-ordered exponential [9], $\mathbf{H}_0$ is the part of the system Hamiltonian that cannot be controlled, $\mathbf{R}$ is the relaxation operator, $\boldsymbol{\rho}_0$ is the initial state vector and $J_{\mathrm{RF}}$ is a penalty functional enforcing physical, instrumental and legal constraints on the radiofrequency or microwave power in the control channels.

Once the controllable part of the Hamiltonian is parameterized, the variation $\delta J/\delta \mathbf{H}_1$ becomes a gradient $\nabla J$ in the parameter space and the process of maximizing $J$ becomes an instance of the very well researched non-linear optimization problem for a continuous function [10,11]. Of the major families of derivative based optimization methods, gradient descent [8], conjugate gradients [12], and quasi-Newton methods [13,14] have all been explored in the optimal control context. In this communication we evaluate the performance of the Newton-Raphson method for optimal control of spin dynamics and propose solutions to the associated numerical problems of Hessian regularization and matrix exponential recycling.

**Newton-Raphson version of the GRAPE algorithm**

The gradient ascent pulse engineering (GRAPE) method [8,13] proceeds by splitting the Hamiltonian into the uncontrollable part $\mathbf{H}_0$ and a number of control operators $\mathbf{H}_k$ with time-dependent coefficients $c_k(t)$

$$\mathbf{H}(t) = \mathbf{H}_0 + \sum_{k=1}^{K} c_k(t)\mathbf{H}_k \qquad (2)$$

The control coefficients are then discretized on a finite grid of time points (usually with a uniform step $\Delta t$) to obtain control sequences in which individual elements $c_k(t_n) = c_{k,n}$ are treated as continuous parameters

$$c_k(t) \quad \to \quad \mathbf{c}_k = \begin{bmatrix} c_k(t_1) & c_k(t_2) & \cdots & c_k(t_N) \end{bmatrix}, \quad t_1 < t_2 < \ldots < t_N \qquad (3)$$

and the local optimality condition is defined as $\nabla J = 0$ with a negative definite Hessian matrix. With a piecewise-constant Hamiltonian, the time-ordered exponential in Equation (1) becomes

$$\exp_{(O)}\left[-i\int_0^T \left(\mathbf{H}_0 + \mathbf{H}_1(t) + i\mathbf{R}\right)dt\right] = \prod_n \exp\left[-i\left(\mathbf{H}_0 + \sum_{k=1}^K c_{k,n}\mathbf{H}_k + i\mathbf{R}\right)\Delta t\right] \quad (4)$$

with a time-ordered product, and the equation itself acquires the following form:

$$J(\mathbf{c}_1,\ldots,\mathbf{c}_K) = \mathrm{Re}\langle\boldsymbol{\delta}|\mathbf{P}_N\ldots\mathbf{P}_2\mathbf{P}_1|\boldsymbol{\rho}_0\rangle - J_{\mathrm{RF}}(\mathbf{c}_1,\ldots,\mathbf{c}_K)$$

$$\mathbf{P}_n = \exp\left[-i\left(\mathbf{H}_0 + \sum_{k=1}^K c_{k,n}\mathbf{H}_k + i\mathbf{R}\right)\Delta t\right] \quad (5)$$

where the $k$ index runs over the control channels and the $n$ index runs over the time steps. A particular strength of the GRAPE method is that the gradient of the fidelity functional

$$\frac{\partial J}{\partial c_{k,n}} = \mathrm{Re}\langle\boldsymbol{\delta}|\mathbf{P}_N\cdots\mathbf{P}_{n+1}\frac{\partial \mathbf{P}_n}{\partial c_{k,n}}\mathbf{P}_{n-1}\cdots\mathbf{P}_1|\boldsymbol{\rho}_0\rangle \quad (6)$$

has the same numerical complexity scaling as the fidelity functional itself [8]. In our previous communication [15] we have pointed out that this also applies to the Hessian of $J$:

$$\frac{\partial^2 J}{\partial c_{j,n}\partial c_{k,n}} = \mathrm{Re}\langle\boldsymbol{\delta}|\mathbf{P}_N\cdots\mathbf{P}_{n+1}\frac{\partial^2 \mathbf{P}_n}{\partial c_{j,n}\partial c_{k,n}}\mathbf{P}_{n-1}\cdots\mathbf{P}_1|\boldsymbol{\rho}_0\rangle \quad (7)$$

$$\frac{\partial^2 J}{\partial c_{j,n}\partial c_{k,m}} = \mathrm{Re}\langle\boldsymbol{\delta}|\mathbf{P}_N\cdots\mathbf{P}_{n+1}\frac{\partial \mathbf{P}_n}{\partial c_{j,n}}\mathbf{P}_{n-1}\cdots\mathbf{P}_{m+1}\frac{\partial \mathbf{P}_m}{\partial c_{k,m}}\mathbf{P}_{m-1}\cdots\mathbf{P}_1|\boldsymbol{\rho}_0\rangle \quad (8)$$

and noted that this situation is highly unusual in non-linear optimization theory – Hessians are normally so expensive that a significant body of work exists on the subject of avoiding their calculation and recovering second derivative information in an approximate way from the gradient history [10,11,16,17]. The recent quasi-Newton BFGS-GRAPE algorithm [13] is an example of such approach. The fact that the Hessian is cheap suggests that Newton-Raphson type algorithms [10,11] with the control sequence update rule at step $s$

$$\mathbf{c}^{(s+1)} = \mathbf{c}^{(s)} - \left[\nabla^2 J^{(s)}\right]^{-1}\nabla J^{(s)} \quad (9)$$

formulated in terms of the gradient $\nabla J$ and the Hessian $\nabla^2 J$ of the fidelity functional $J(\mathbf{c})$ with a suitable line search procedure are a natural next step.

In principle, Equations (1)-(9) define the algorithm completely. However, three logistical problems present themselves that must be solved before the method becomes useful in practice: (a) efficient calculation of matrix exponential derivatives, (b) exponential derivative recycling between Equations (6)-(8) that is required for efficient scaling, and (c) regularization of the Hessian matrix in Equation (9) that is needed to avoid the numerical difficulties [10,11] associated with its inverse. Solution to the first problem is known – our experience indicates that the auxiliary matrix method [15,18] based on the following block matrix relation

$$\begin{pmatrix} \mathbf{P}_n & \dfrac{\partial \mathbf{P}_n}{\partial c_{i,n}} & \mathbf{A}_{ij,n} \\ \mathbf{0} & \mathbf{P}_n & \dfrac{\partial \mathbf{P}_n}{\partial c_{j,n}} \\ \mathbf{0} & \mathbf{0} & \mathbf{P}_n \end{pmatrix} = \exp\left[-i \begin{pmatrix} \mathbf{H}(t) & \mathbf{H}_i & \mathbf{0} \\ \mathbf{0} & \mathbf{H}(t) & \mathbf{H}_j \\ \mathbf{0} & \mathbf{0} & \mathbf{H}(t) \end{pmatrix} \Delta t \right], \quad \dfrac{\partial^2 \mathbf{P}_n}{\partial c_{i,n} \partial c_{j,n}} = \mathbf{A}_{ij,n} + \mathbf{A}_{ji,n} \quad (10)$$

works very well – a 3×3 block matrix is used to compute the second derivative and a 2×2 block matrix when only the first derivative is required [15]. The problems of derivative recycling and Hessian regularization, however, are non-trivial – they are dealt with in the next three sections.

**Hessian calculation benchmarks**

Our numerical implementations of gradient and Hessian calculations [19] are parallelized with respect to the number of time slices in Equation (7). Gradient calculation uses 2×2 augmented exponentials [15] and Hessian calculation uses the 3×3 augmented exponentials shown in Equation (10). The Hessian function has a further parallel loop when calculating $n \neq m$ blocks in Equation (8). At that stage, the first derivatives have already been calculated when solving Equation (7) with Equation (10) from the block superdiagonal, they are recycled.

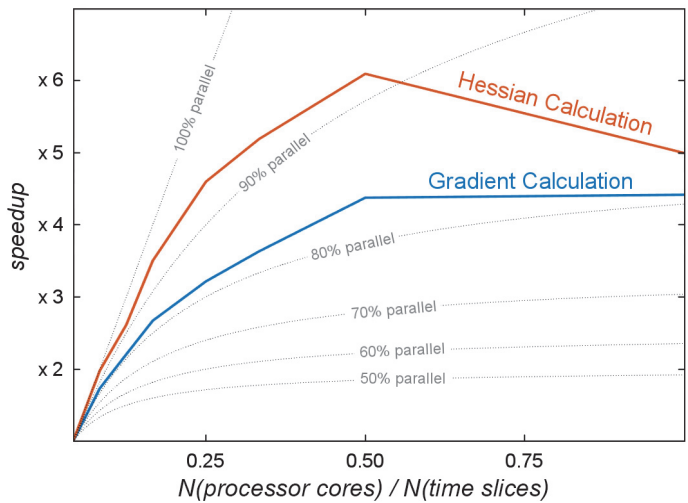

*Figure 1.* Amdahl's law [20] parallelisation efficiency analysis for the Hessian calculation compared to the gradient calculation within Spinach implementation of GRAPE [13,19]. The optimal control problem involves 24 time slices and 6 control channels, yielding a fidelity functional gradient with 144 elements and a 144×144 Hessian.

Parallelisation efficiency analysis is given in Figure 1. The scaling is independent of the number of time slices in the control sequence and the parallelism is good all the way to the number of CPU cores being half the number of time slices (over which the parallel loop is running). Given the same computing resources, a Hessian calculation takes approximately ten times longer than a gradient calculation. Because propagator derivatives are recycled during Hessian calculation, significant efficiency gains may be made by optimising their storage and indexing – this is the subject of the following section.

**Matrix function recycling**

Repeated evaluation of expensive matrix functions of the same argument is a common problem in quantum dynamics simulations. In our context this problem is evident in Equations (6)-(8) where exponentials of the same matrix are expected to occur multiple times. Storing all previously encountered arguments and running a database search at each call would be inconvenient and inefficient. Split operator methods [21] that could avoid the expensive matrix exponentiation step are not applicable to spin dynamics because there is no notion of coordinate or momentum, and accurate wavefunction phase is required at all times. Krylov exponentiation may occasionally be used for propagation [22] when $j \neq k$ in Equation (7), but the N×K diagonal Hessian elements in Equation (7) still involve derivatives of matrix exponentials explicitly. In this section we propose a method that, subject to sufficient storage being available, enables convenient and efficient recycling of expensive matrix functions.

Simply looking up a matrix in a database of previously encountered ones is out of question for the following reasons. On modern computer architectures a memory retrieval and comparison operation takes at least one CPU clock cycle. The time cost of looking up a given matrix in a sorted list of previously encountered ones is therefore at least $O(N_{NZ} \log N_M)$ clocks [23], where $N_{NZ}$ is the number of non-zero elements in the matrix, and $N_M$ is the number of matrices in the database. The worst-case sorting cost for the database of previously encountered matrices is $O(N_{NZ} N_M \log N_M)$ clocks [23], which is unacceptable because the number of non-zeros in commonly encountered matrices can be in the millions.

The standard solution from database theory is to use a hash table [24]. This being a physical sciences journal, a detailed exposition is perhaps warranted. A cryptographic *hash function* is a function that accepts an input (called *message*) of any length and produces an output (called *digest*) with the following properties [25]:

1. The complexity of computing the digest is linear with the size of the message.

2. A single-bit modification in the message is *almost certain* to change its digest.

3. Two randomly selected messages are *almost certain* to have different digests.

The first property offers a solution to our lookup cost problem: the complexity of computing the hash is $O(N_{NZ})$ – negligible compared to the cost of expensive matrix factorizations [26] and significantly smaller than the direct sorting and lookup costs discussed above. Hashing a matrix is a straightforward procedure: for full matrices, the array is typecast (in-place, to avoid making a memory copy) into UINT8 and fed into a hashing engine. For sparse matrices, the index array, the value array and the array of matrix dimensions are typecast into UINT8, concatenated and hashed as a single message. The cost of sorting the hash table is $O(N_M \log N_M)$ and the cost of looking up a digest in the sorted list is $O(\log N_M)$ [23]. This reduces the total cost of the matrix lookup to $O(N_{NZ}) + O(N_M \log N_M) + O(\log N_M)$. In situations where matrix operation caching is necessary, $N_{NZ} \gg N_M \log N_M \gg \log N_M$ (this may also be viewed as the condition under which it is sensible to use matrix operation caching) – the overall asymptotic cost of hash table matrix lookup and all the associated housekeeping is therefore $O(N_{NZ})$ clocks. It should be noted that extremely efficient hashing hardware has recently become available as a side effect of the emergence of cryptocurrencies [27].

The second and the third properties provide collision safety assurances: the definition of *almost certain* in this context is that one needs to calculate $2^{N/2}$ hash values (where $N$ is the number of bits in the digest) to have a 50% probability of seeing a hash collision [28]. Even with basic 128-bit hash functions, such as MD5, this is a vanishingly rare event in the physical sciences context: a useful rule of thumb in modern physics is that any probability smaller than that of the researcher committing suicide (approximately $9.7 \times 10^{-5}$ per year in the UK in 2013 [29]) is negligible. Caching algorithms based on MD5 and SHA hash functions fall below that measure by many orders of magnitude – for our purposes they are safe. If absolute certainty is required, an additional step of comparing the matrices element-by-element may be added, at the cost of extra storage, but without any changes to the asymptotic $O(N_{NZ})$ complexity scaling estimate.

The benefit derived from the caching procedure has the same caveats as the well-researched algorithms for caching disk access [30] – when the same sectors are requested repeatedly, the benefit is large, but for random access the cache can actually make the process slower. Matrix function caching should therefore only be used in situations where repeated requests for expensive functions of the same argument are likely – that situation is thankfully very frequent. The hashing and caching procedure may also be viewed as a generalization of the concept of a look-up grid, but without the possibility of interpolation. Implementing an interpolated look-up grid would of course be impractical due to the large dimension of the interpolation problem space.

In the quantum dynamics context, the increase in performance resulting from using matrix exponential caching is illustrated in Figure 2. The CN2D NMR pulse sequence [31] is designed to correlate $^{14}$N and $^{13}$C NMR signals under magic angle spinning conditions. It contains multiple periods that have identical Hamiltonians – the caching algorithm identifies those automatically and avoids their recalculation.

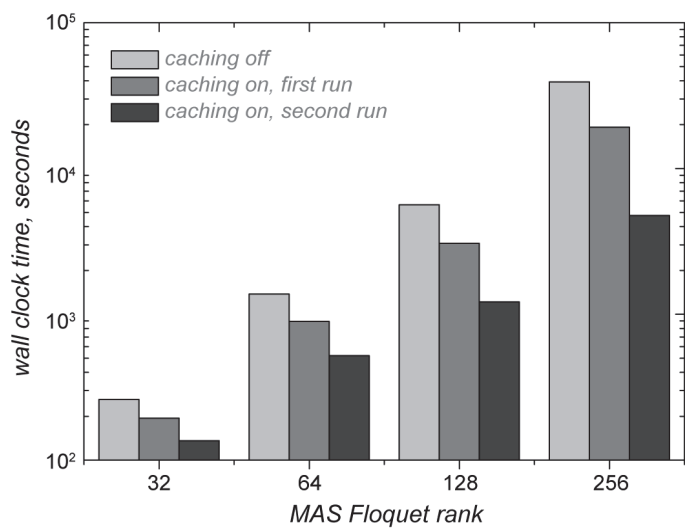

*Figure 2. Wall clock time consumed by the simulation of the CN2D solid state NMR experiment [31] for a $^{14}$N-$^{13}$C spin pair in glycine with different matrix exponential caching settings in Spinach [19]. Light grey columns correspond to running with matrix caching switched off, medium grey columns are for runs with the caching switched on and a cache that is empty at the start of the simulation. Dark grey columns correspond to simulations where all required matrix exponentials are already present in the cache. Details of the NMR pulse sequence and the spin Hamiltonians involved are given in Reference [31].*

It is in principle possible to hand-code this simulation in such a way as to avoid repeated calls to expensive functions by manually identifying the time intervals that have identical Hamiltonians. Such an approach would not, however, be scalable to more complicated experiments and to highly general and automated simulation systems, such as *Spinach* [19].

Cache destination can be a file system or a key-value store of any type – a solid state disk was used in this work. In practice, the latency of the cache storage device means that for small matrices it may be faster to recalculate the function. In our practical experience, the caching procedure becomes beneficial once the dimension exceeds 512. Because matrix exponentiation dominates the numerical cost of optimal control simulations, a beneficial side effect is a rapid restart capability for the GRAPE algorithm – the calculation can re-trace its steps quickly.

**Hessian regularization**

Newton-Raphson and quasi-Newton methods (minimisation is assumed here) rely on the necessary conditions for Taylor's theorem [32,33] and use a local quadratic approximation:

$$J(\mathbf{c}+\Delta\mathbf{c}) \approx J(\mathbf{c}) + \langle \nabla J(\mathbf{c}) | \Delta\mathbf{c} \rangle + \tfrac{1}{2} \langle \Delta\mathbf{c} | \nabla^2 J(\mathbf{c}) | \Delta\mathbf{c} \rangle \tag{11}$$

The first order necessary condition requires any minimiser $\tilde{\mathbf{c}}$ to be a stationary point

$$\nabla J(\tilde{\mathbf{c}}) = 0 \tag{12}$$

Imposing this condition on Equation (11) gives the control sequence update rule from Equation (9). The second order necessary condition is that the Hessian $\nabla^2 J$ should be positive definite at $\tilde{\mathbf{c}}$. This is also evident from Equation (9), in which a negative Hessian eigenvalue would result in a step being performed up, rather than down, the corresponding gradient direction.

A significant problem is that, far away from a minimiser, the Hessian is not actually expected to be positive definite. Small Hessian eigenvalues are also problematic because they result in overly long steps that can be detrimental because most fidelity functionals are not actually quadratic. A significant amount of research has gone into modifying the Hessian in such a way as to avoid these undesired behaviours [34-44].

One fairly cheap way to work around an indefinite Hessian is to attempt Cholesky factorization, which exists for any invertible positive definite matrix [45]:

$$\nabla^2 J = \mathbf{L}\mathbf{L}^\mathrm{T} \quad \Rightarrow \quad \left[\nabla^2 J\right]^{-1} = \mathbf{L}^{-1,\mathrm{T}}\mathbf{L}^{-1} \tag{13}$$

where $\mathbf{L}$ is a lower triangular matrix and $\mathbf{L}^\mathrm{T}$ is its transpose. If this fails, an identity matrix may be used as a substitute for the Hessian, effectively reverting to a gradient descent step for any iterations that produce an indefinite Hessian [46]. The problem with this approach is that indefinite Hessians become more common as the dimension of the problem increases, making the minimizer spend most of the time in the gradient descent mode and destroying any advantage of the second-order method over simple gradient descent – we do not recommend this technique.

A more sophisticated workaround is to use the eigenvalue shifting method (*aka* trust region method, TRM), suggested by the origins of the Levenberg-Marquardt algorithm [47,48] and us-ing the Cholesky decomposition [49] on a Hessian with a multiple of the unit matrix added [10,34,37,40,42,50]:

$$\nabla^2 J + \sigma \mathbf{1} = \mathbf{L}\mathbf{L}^T, \quad \sigma \geq 0 \tag{14}$$

The choice of $\sigma$ is made to produce a positive definite Hessian, with the trial value

$$\sigma = \begin{cases} \left\|\nabla^2 J\right\|_F - \min\left[\nabla^2 J\right]_{ii} & \text{if} \quad \min\left[\nabla^2 J\right]_{ii} < 0 \\ \left\|\nabla^2 J\right\|_F & \text{if} \quad \min\left[\nabla^2 J\right]_{ii} \geq 0 \end{cases} \tag{15}$$

chosen in that way because the Frobenius norm of the Hessian is an upper bound on the largest absolute eigenvalue. The value of $\sigma$ is increased iteratively until the Cholesky decomposition succeeds [11] and the inverse Hessian may be obtained. Alternatively, Hessian eigenvalues may be computed explicitly [35] and a precise estimate made for the value of $\sigma$ in Equation (14):

$$\left[\nabla^2 J\right] = \mathbf{Q}\mathbf{\Lambda}\mathbf{Q}^{-1}, \quad \sigma = \max(0, \delta - \min(\mathbf{\Lambda}_{ii}))$$
$$\left[\nabla^2 J\right]_{\text{reg}} = \mathbf{Q}(\mathbf{\Lambda} + \sigma\mathbf{1})\mathbf{Q}^{-1} \tag{16}$$

where $\mathbf{\Lambda}$ is a diagonal matrix containing the eigenvalues of $\nabla^2 J$ and $\mathbf{Q}$ is the matrix with columns made up of corresponding eigenvectors. The user-specified positive value of $\delta$ is included to make the Hessian positive definite. The primary problem with this method is that, for poorly conditioned Hessian matrices, the regularization procedure destroys much of the curvature information and the technique effectively becomes a combination of Newton-Raphson method and gradient descent. Based on our practical performance evaluation, it is also not recommended here: formally, Equations (14) and (16) do solve the step direction problem, but the convergence rate we have seen in practice was not superior to gradient descent.

The method that was found to perform best in our practical testing is known as rational function optimization (RFO) [36,41,43]. It replaces the Taylor expansion with a Padé approximant [51]:

$$\Delta J = \frac{\langle \nabla J | \mathbf{c} \rangle + \tfrac{1}{2}\langle \mathbf{c} | \nabla^2 J | \mathbf{c} \rangle}{1 + \langle \mathbf{c} | \mathbf{S} | \mathbf{c} \rangle} \tag{17}$$

This preserves the derivative information in Equation (11), leaving the necessary conditions unchanged, because the derivatives of $\left[1 + \langle \mathbf{c} | \mathbf{S} | \mathbf{c} \rangle\right]^{-1}$ give contributions only through higher orders in $\mathbf{c}$. The nature of the rational function means that the asymptotes of $\Delta J$ and its gradients remain finite for $\|\mathbf{c}\|_\infty \to \pm\infty$, determined by the Hessian and the symmetric scaling matrix $\mathbf{S}$. The first order necessary condition for Equation (12) gives the following eigenvalue equation:

$$\begin{pmatrix} \nabla^2 J & \nabla J \\ \nabla J^T & 0 \end{pmatrix} \begin{pmatrix} \mathbf{c} \\ 1 \end{pmatrix} = 2\Delta J \begin{pmatrix} \mathbf{S} & 0 \\ 0 & 1 \end{pmatrix} \begin{pmatrix} \mathbf{c} \\ 1 \end{pmatrix} \tag{18}$$

Choosing a uniform scaling matrix [43] $\mathbf{S} = \alpha^{-2}\mathbf{1}$, where $0 < \alpha \leq 1$, reduces this equation to

$$\begin{pmatrix} \alpha^2 \nabla^2 J & \alpha \nabla J \\ \alpha \nabla J^T & 0 \end{pmatrix} \begin{pmatrix} \mathbf{c}/\alpha \\ 1 \end{pmatrix} = 2\Delta J \begin{pmatrix} \mathbf{c}/\alpha \\ 1 \end{pmatrix} \tag{19}$$

Rational function optimization proceeds in a similar way to eigenvalue shifting methods described above, except the shifting is applied to the augmented Hessian:

$$\left[\nabla^2 J\right]^{aug} = \begin{pmatrix} \alpha^2 \nabla^2 J & \alpha \nabla J \\ \alpha \nabla J^T & \mathbf{0} \end{pmatrix} = \mathbf{Q}\boldsymbol{\Lambda}\mathbf{Q}^{-1}, \quad \sigma = \max(0, -\min(\boldsymbol{\Lambda}_{ii}))$$
$$\left[\nabla^2 J\right]^{aug}_{reg} = \frac{1}{\alpha^2}\mathbf{Q}(\boldsymbol{\Lambda} + \sigma\mathbf{1})\mathbf{Q}^{-1} \quad (20)$$

The top left corner block of the regularized augmented Hessian is then used for the Newton-Raphson step [36,41,43] in Equation (9). Our practical experience with the scaling constant $\alpha$ indicates that it should be allowed to vary, with $\alpha = 1$ when the Hessian is well-conditioned and a value that is reduced until the condition number becomes acceptable, for example:

$$\alpha_{r+1} = \phi \alpha_r \quad \text{while} \quad \frac{\min(\boldsymbol{\Lambda}_{ii})}{\max(\boldsymbol{\Lambda}_{ii})} > \frac{1}{\sqrt[n]{\varepsilon}} \quad (21)$$

where $\varepsilon$ is machine precision and $\alpha_0 = 1$. The factor $0 < \phi < 1$ is used to iteratively decrease the condition number of the Hessian – this is the method used to condition the Hessian in the examples presented below. It should be noted that a value $\phi > 1$ may be used to increase the condition number of the Hessian when it is very small and the inequality of Equation (21) is reversed. The $n$ root of $\varepsilon$ appearing in Equation (21) as the strict limit for a line search method using polynomial interpolation of degree $n$.

In our practical experience, the end users cannot unfortunately be relied upon to scale their problem well – quantum mechanics is rich in situations that produce very small values of Hessian and gradient elements. We therefore propose choosing $\alpha_0$ to be the value that would shift the smallest eigenvalues to be above 1, giving a similar effect to choosing $\delta = 1$ in Equation (16). When $\alpha_0 = \sqrt{1/|\min(\boldsymbol{\Lambda}_{ii})|} = \sqrt{1/\lambda_{min}}$ choice is made, the augmented Hessian in Equation (20) becomes

$$\left[\nabla^2 J\right]^{aug} = \begin{pmatrix} \frac{1}{\lambda_{min}}\nabla^2 J & \frac{1}{\sqrt{\lambda_{min}}}\nabla J \\ \frac{1}{\sqrt{\lambda_{min}}}\nabla J^T & \mathbf{0} \end{pmatrix} = \mathbf{Q}\boldsymbol{\Lambda}\mathbf{Q}^{-1}, \quad \sigma = \max(0, -\min(\boldsymbol{\Lambda}_{ii}))$$
$$\left[\nabla^2 J\right]^{aug}_{reg} = \lambda_{min}\mathbf{Q}(\boldsymbol{\Lambda} + \sigma\mathbf{1})\mathbf{Q}^{-1} \quad (22)$$

Our testing indicates that this combination of using initial scaling and then accepting large condition numbers for the Hessian allows the Newton-Raphson method to avoid getting stuck at inflection points. The condition number would grow to a large value around the inflection point and then shrink back when the point has been avoided. Due to the tendency of the Hessian to increase condition number as it approaches the minimizer, machine precision could eventually become a limit to this type of conditioning. To avoid a slowdown at the final stages of the optimization, an upper bound is placed on the $\alpha$ parameter in Equation (21) – this guarantees that the terminal convergence is always quadratic.

In practice, at each optimization step the function code attempts to compute the Cholesky decomposition of Equation (13). If that is successful then no regularization is needed, otherwise the function proceeds to regularize with the methods described above. Once the descent direction and the initial step length are obtained from Equation (9) it is advantageous to perform a line search procedure. We adapted the version with a bracketing phase that obeys strong Wolfe con-ditions [52,53] followed by a sectioning phase with cubic interpolation. The initial step length is always equal to that of Equation (9) to ensure that the convergence is at least quadratic [11].

**Derivatives of penalty functionals**

The inevitable limitations of physical hardware on maximum amplitude and minimum switching time necessitate the addition of the penalty term that would enforce those limitations – that is the role of the $J_{RF}$ functional in Equation (1). Instrumental limitations are rarely hard and it is therefore reasonable to implement them as penalties rather than hard bounds. For Newton type optimizations to work, second derivatives of $J_{RF}$ with respect to the control sequence are required, they are given in Table 1 for the three most common penalty functional types.

**Table 1.** Common GRAPE penalty functionals with their first and second derivatives.

| Functional type | Penalty | Penalty gradient | Penalty Hessian |
|---|---|---|---|
| weighted norm square | $J_{RF} = \sum_k w_k c_k^2$ | $[\nabla J_{RF}]_n = 2 w_n c_n$ | $[\nabla^2 J_{RF}]_{nm} = 2 w_n \delta_{nm}$ |
| weighted derivative norm square | $J_{RF} = \sum_k w_k [\mathbf{Dc}]_k^2$ | $[\nabla J_{RF}]_n = 2 \sum_k w_k [\mathbf{Dc}]_k D_{kn}$ | $[\nabla^2 J_{RF}]_{nm} = 2 \sum_k w_k D_{kn} D_{km}$ |
| weighted spillout norm square | $J_{RF} = \sum_k w_k (c_k - u_k)^2 h_{c_n > u_n} + \sum_k w_k (c_k - l_k)^2 h_{c_n < l_n}$ | $[\nabla J_{RF}]_n = 2 w_n (c_n - u_n) h_{c_n > u_n} + 2 w_n (c_n - l_n) h_{c_n < l_n}$ | $[\nabla^2 J_{RF}]_{nm} = 2 w_n \delta_{nm} h_{c_n > u_n} + 2 w_n \delta_{nm} h_{c_n < l_n}$ |

In Table 1, $\mathbf{c}$ is the control sequence vector, $\mathbf{w}$ is the weight vector, $\mathbf{u}$ is the upper bound vector, $\mathbf{l}$ is the lower bound vector and $\mathbf{D}$ is the differentiation matrix of a suitable type and order, or any other appropriate transformation matrix. The weighted norm square spillout penalty is designed to only apply to the parts of the control sequence that fall outside the bounds defined by $\mathbf{u}$ and $\mathbf{l}$ vectors. The meaning of the corresponding Heaviside and delta functions is:

$$h_{c_n > u_n} = \begin{cases} 1 & \text{if } c_n > u_n \\ 0 & \text{otherwise} \end{cases} \qquad h_{c_n < l_n} = \begin{cases} 1 & \text{if } c_n < l_n \\ 0 & \text{otherwise} \end{cases} \qquad \delta_{nm} = \begin{cases} 1 & \text{if } n = m \\ 0 & \text{otherwise} \end{cases} \qquad (23)$$

The principal advantage of the three functionals listed above over the multitude of possible alternatives is in the fact that the corresponding derivatives are cheap to compute. All three are implemented in the penalty module of *Spinach* [19].

**Performance analysis**

Several performance metrics are possible for optimal control algorithm benchmarking: wall clock time to a given fidelity, iteration count to a given fidelity, function evaluation count to a given fidelity, *etc.* Our choice of metric is dictated by the motivation of this work: to accelerate waveform optimization in time-critical scenarios, such as clinical MRI, assuming that a parallel computer with the number of CPU cores equal to at least half the number of time slices is available. That computer needs to run Equations (5), (6) and (7), of which only Equation (5) is not highly parallel because it involves propagating the system from $|\rho_0\rangle$ to $|\rho(T)\rangle$ and keeping all intermediate trajectory points that are re-used in Equations (6) and (7). Trajectory generation is a serial process because the next trajectory point depends on the previous one. In contrast, Equations (6) and (7) have the same asymptotic cost as Equation (5), but are easy to parallelise – at the limit, every element of the gradient array and the Hessian array may be sent to a separate core. The wall clock time on large parallel computers is therefore determined by the number of system trajectory calculations because the generation of the system trajectory is the only serial step. At each GRAPE iteration, only one Hessian calculation or update is made, but further trajectory calculations may be required within the line search stage that uses a gradient-based bracketing and sectioning line search with cubic interpolation and strong Wolfe conditions [11]. We shall therefore use both the trajectory calculation count and the optimization algorithm iteration count as the time variable in the benchmarks presented below.

The first test system is an H–C–F group in a 9.4 Tesla magnet with $^1$H isotope for hydrogen, $^{13}$C isotope for carbon and $^{19}$F isotope for fluorine, with the $^1$H-$^{13}$C $J$-coupling of 140 Hz, $^{13}$C-$^{19}$F $J$-coupling of –160 Hz and all three signals assumed to be on resonance with the transmitters on the corresponding NMR spectrometer channels. A six-channel (H$_X$, H$_Y$, C$_X$, C$_Y$, F$_X$, F$_Y$) shaped pulse with a duration of 100 ms, a quadratic penalty for excursions outside the 10 kHz power envelope and 50 time discretisation points was optimized to perform longitudinal magnetization transfer from $^1$H to $^{19}$F. This system was chosen because very high terminal fidelities are achievable with the parameters described above – it is a good test of terminal convergence behaviour for quadratic optimization algorithms in finite precision arithmetic.

Fidelity functional optimization profiles using TRM and RFO regularization with Newton-Raphson method are compared in Figure 3 with the previously leading method for this class of optimal control problems – the BFGS quasi-Newton method [13,14]. It is clear that Hessian regularization is always advantageous and that RFO regularization performs better than TRM. It is also clear that the exactly quadratic optimization method beats the asymptotically quadratic one by a considerable margin.

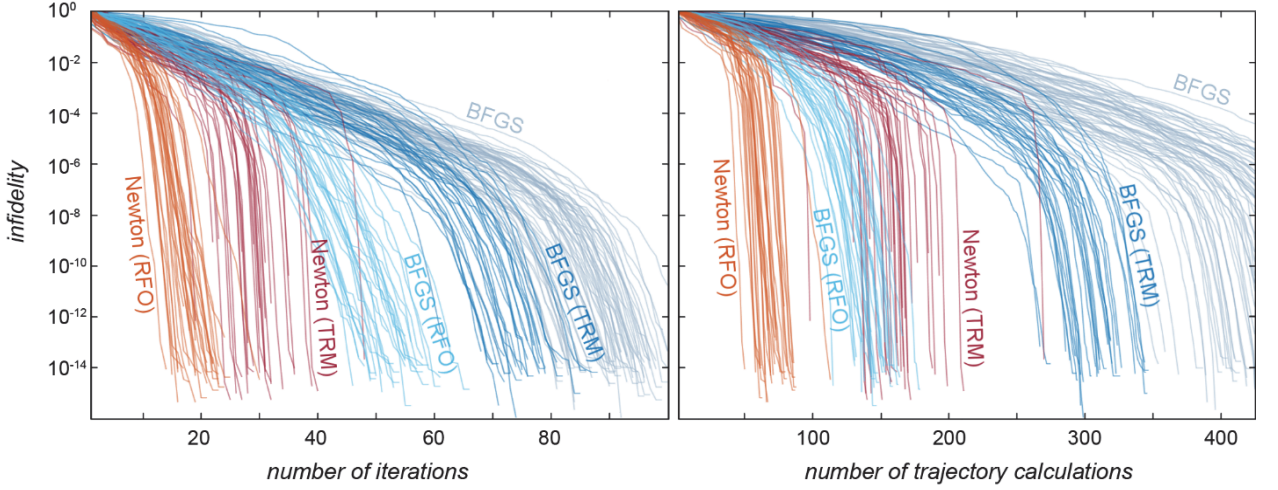

***Figure 3.*** *Convergence profiles for the state transfer within the $^1$H–$^{13}$C–$^{19}$F three-spin system described in the main text for the BFGS quasi-Newton method and the Newton-Raphson method using TRM or RFO Hessian regularization techniques. The same line search method in the predicted descent direction was used in all cases.*

The primary reason for the better performance of RFO is the regular asymptotic behaviour of Equation (17) and its derivatives compared to the simple Taylor expansion in Equation (11). The Cholesky decomposition method in TRM, although able to force a positive definite Hessian, offers little influence over its condition number. This leads to poorly balanced step directions in which the smallest eigenvalues dominate and much of the local curvature information is lost in the regularization process, leading to high costs at the subsequent line search stage.

The second test case involves a state transfer from longitudinal polarization into a two-spin singlet state, while allowing for up to 20% miscalibration of the control channel power level. The spin system contains two $^{13}$C spins in a 14.1 Tesla magnet with chemical shifts of 0.00 and 0.25 ppm and a *J*-coupling of 60 Hz. The system is prepared in the $\hat{C}_Z^{(1)} + \hat{C}_Z^{(2)}$ state and a two-channel control sequence on $\hat{C}_X^{(1)} + \hat{C}_X^{(2)}$ and $\hat{C}_Y^{(1)} + \hat{C}_Y^{(2)}$ control operators with 50 time discretization points, the nominal power of 60 Hz and the duration of 50 milliseconds is optimized simultaneously for ten different power levels spaced equally between 80% and 120% of the nominal power. Weighted norm squared penalty functional was used (Table 1) with equal weights for all time points. The infidelity measure in Figure 4 refers to the distance from the best possible magnetization transfer fidelity for the system in question. Trajectory analysis diagrams and the optimal control sequences are presented in Figure 5.

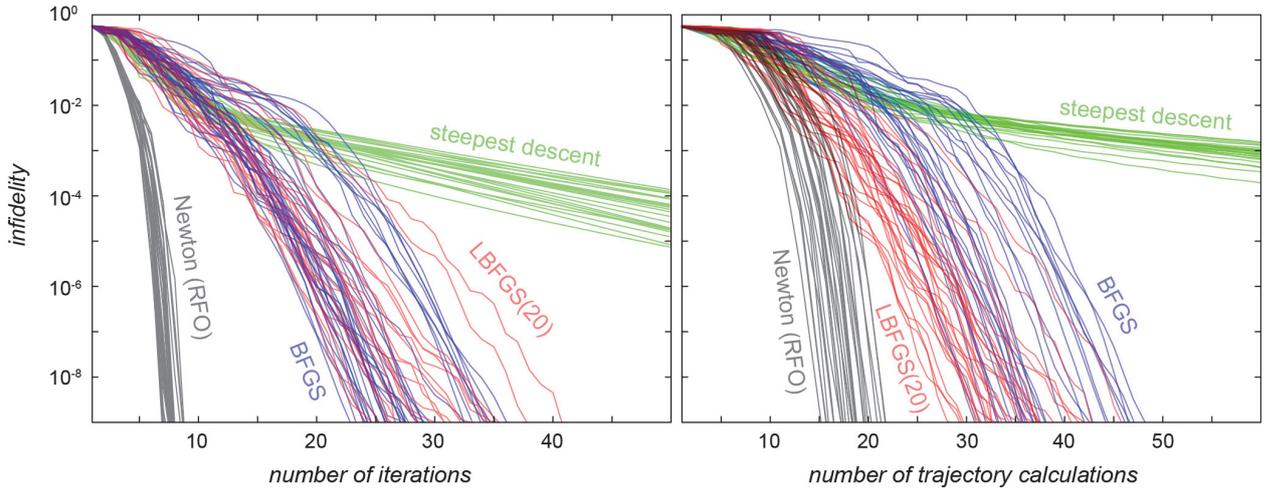

*Figure 4.* Convergence profiles for the transfer of longitudinal magnetisation into the singlet state for the two-spin system described in the main text. The same line search method in the predicted descent direction was used in all cases. Memory time for LBFGS was set to 20 gradients.

Figure 4 illustrates the common wisdom that gradient descent and its variations are only good up to a point and optimal control problems that require high fidelities must use methods that have, exactly or asymptotically, quadratic convergence behaviour [13]. It is also clear that the RFO regularised Newton-Raphson method described above is superior to quasi-Newton methods and vastly superior to gradient descent. This relationship between the three classes of methods is to be expected [11] – our claim to novelty here is to demonstrate that the Newton method is actually affordable for quantum optimal control because the off-diagonal elements in Equation (8) can re-use propagator derivatives from Equation (7) and the diagonal elements have the same asymptotic cost as Equation (6). There is no longer any excuse for using steepest descent.

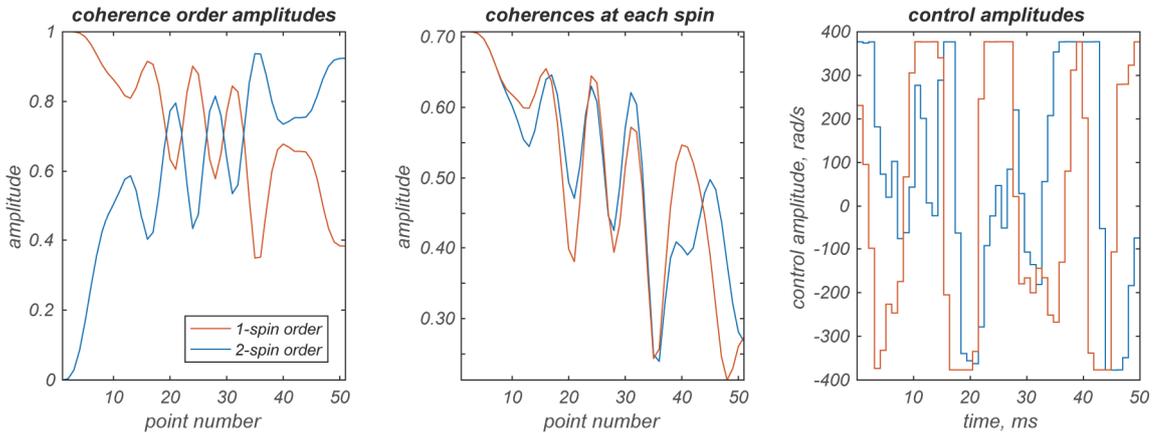

*Figure 5.* Spin system trajectory analysis and the control amplitudes for the state transfer from longitudinal magnetization into the singlet state for the two-spin system described in the main text. Trajectory analysis methods used to obtain the left and the middle panel are described in [54].

While a performance comparison with other methods belonging to the GRAPE family is possible and rather favourable (Figures 3 and 4), comparing the Newton GRAPE technique directly to the Krotov family of optimal control algorithms [7,14] would not be appropriate here because of the very different iteration structure of those methods. Krotov type algorithms update the controls sequentially within the time propagation loop. This means that there is, strictly speaking, no gradient or Hessian of the fidelity functional occurring anywhere within the Krotov iteration. Second-order information may still be used to accelerate convergence [14], but any direct performance comparisons would be open to criticism because "iteration" means different things within Krotov and GRAPE frameworks. Similar considerations apply to more specialised second-order methods, such as matrix-free Krylov-Newton algorithms [55,56] that use Crank-Nicholson time propagation instead of the exponential propagation used in this paper. Having witnessed a few benchmark wars in optimal control already, we would rather not start another one in a situation where comparisons might not be meaningful.

All methods described in this communication are implemented in versions 1.8 and later of *Spinach* library [19]. Both test systems described in this section are in the example set. *Spinach* console logs giving complete spin system and algorithm setting details for the simulations described above are included in the supplementary information [54].

**Conclusions**

The GRAPE algorithm for optimal control of quantum systems is highly unusual in that the Hessian of the fidelity functional has the same asymptotic computational cost as the gradient when due care is taken to recycle the intermediate results for the matrix exponential derivatives. This makes the usually prohibitively expensive, but very efficient, Newton-Raphson optimization algorithm affordable. After all technical problems with Hessian regularization, line search and matrix exponential recycling are addressed, the result is an optimal control algorithm that converges faster than BFGS-GRAPE (which is the current leader) and much faster than all previous GRAPE implementations. It is recommended particularly in time-constrained situations, such as magnetic resonance imaging, due faster convergence and greater code parallelisation opportunities compared to other quantum control algorithms in the GRAPE family.

**Acknowledgements**

We are grateful to Sophie Schirmer, Stefan Stoll, Thomas Schulte-Herbrüggen and Steffen Glaser for useful discussions. This work was made possible by EPSRC through a grant to IK group (EP/H003789/1) and a CDT studentship to DLG. The European Commission has facilitated this project through a coordination action grant (297861/QUAINT).

**References**

[1] M.F. Dempsey, B. Condon, D.M. Hadley, MRI safety review, Seminars in Ultrasound, CT and MRI, 23 (2002) 392-401.

[2] P.L. Gor'kov, E.Y. Chekmenev, C. Li, M. Cotten, J.J. Buffy, N.J. Traaseth, G. Veglia, W.W. Brey, Using low-E resonators to reduce RF heating in biological samples for static solid-state NMR up to 900 MHz, Journal of Magnetic Resonance, 185 (2007) 77-93.

[3] Y. Iwasa, HTS and NMR/MRI magnets: Unique features, opportunities, and challenges, Physica C: Superconductivity and its Applications, 445–448 (2006) 1088-1094.

[4] S.A. Sarji, B.J.J. Abdullah, G. Kumar, A.H. Tan, P. Narayanan, Failed magnetic resonance imaging examinations due to claustrophobia, Australasian Radiology, 42 (1998) 293-295.

[5] L.S. Pontryagin, V.G. Boltanskii, R.S. Gamkrelidze, E.F. Mishchenko, The mathematical theory of optimal processes, Pergamon, 1964.

[6] S.J. Glaser, U. Boscain, T. Calarco, C.P. Koch, W. Köckenberger, R. Kosloff, I. Kuprov, B. Luy, S. Schirmer, T. Schulte-Herbrüggen, Training Schrödinger's cat: Quantum optimal control, The European Physical Journal D, 69 (2015) 1-24.

[7] Y. Maday, G. Turinici, New formulations of monotonically convergent quantum control algorithms, The Journal of Chemical Physics, 118 (2003) 8191-8196.

[8] N. Khaneja, T. Reiss, C. Kehlet, T. Schulte-Herbruggen, S.J. Glaser, Optimal control of coupled spin dynamics: design of NMR pulse sequences by gradient ascent algorithms, J Magn Reson, 172 (2005) 296-305.

[9] D.J. Tannor, Introduction to quantum mechanics, A Time-Dependent Perspective, 75 (2007).

[10] R. Fletcher, Practical methods of optimization, 2nd ed., Wiley, 1987.

[11] J. Nocedal, S.J. Wright, Numerical optimization, 2nd ed., Springer, 2006.

[12] L. Lasdon, S. Mitter, A. Waren, The conjugate gradient method for optimal control problems, Automatic Control, IEEE Transactions on, 12 (1967) 132-138.

[13] P. de Fouquieres, S.G. Schirmer, S.J. Glaser, I. Kuprov, Second order gradient ascent pulse engineering Journal of Magnetic Resonance, 212 (2011) 412 - 417.

[14] R. Eitan, M. Mundt, D.J. Tannor, Optimal control with accelerated convergence: Combining the Krotov and quasi-Newton methods, Physical Review A, 83 (2011) 053426.

[15] D.L. Goodwin, I. Kuprov, Auxiliary matrix formalism for interaction representation transformations, optimal control, and spin relaxation theories, The Journal of chemical physics, 143 (2015) 084113.

[16] D.C. Liu, J. Nocedal, On the limited memory BFGS method for large scale optimization, Math. Progr., 45 (1989) 503-528.

[17] R.H. Byrd, J. Nocedal, R.B. Schnabel, Representations of quasi-Newton matrices and their use in limited memory methods, Mathematical Programming, 63 (1994) 129-156.

[18] I. Najfeld, T.F. Havel, Derivatives of the Matrix Exponential and Their Computation, Advances in Applied Mathematics, 16 (1995) 321-375.

[19] H.J. Hogben, M. Krzystyniak, G.T. Charnock, P.J. Hore, I. Kuprov, Spinach - a software library for simulation of spin dynamics in large spin systems, J Magn Reson, 208 (2011) 179-194.

[20] G.M. Amdahl, Validity of the single processor approach to achieving large scale computing capabilities, in: Proceedings of the April 18-20, 1967, spring joint computer conference, ACM, 1967, pp. 483-485.

[21] A.D. Bandrauk, H. Shen, Improved exponential split operator method for solving the time-dependent Schrödinger equation, Chemical physics letters, 176 (1991) 428-432.

[22] R.B. Sidje, Expokit: a software package for computing matrix exponentials, ACM Transactions on Mathematical Software (TOMS), 24 (1998) 130-156.

[23] T.H. Cormen, Introduction to algorithms, MIT press, 2009.

[24] W.D. Maurer, T.G. Lewis, Hash table methods, ACM Computing Surveys (CSUR), 7 (1975) 5-19.

[25] R. Sedgewick, Algorithms in Java, Addison-Wesley Professional, 2002.

[26] D.E. Knuth, The art of computer programming, Addison-Wesley, Upper Saddle River, NJ, 2005.

[27] N.T. Courtois, M. Grajek, R. Naik, Optimizing SHA256 in bitcoin mining, in: Cryptography and Security Systems, Springer, 2014, pp. 131-144.

[28] X. Wang, H. Yu, How to break MD5 and other hash functions, in: Advances in Cryptology–EUROCRYPT 2005, Springer, 2005, pp. 19-35.

[29] D. Evans, UK Office for National Statistics - Suicides in the United Kingdom, 2013.

[30] B. Jacob, S. Ng, D. Wang, Memory systems: cache, DRAM, disk, Morgan Kaufmann, 2010.

[31] J.A. Jarvis, I.M. Haies, P.T.F. Williamson, M. Carravetta, An efficient NMR method for the characterisation of 14N sites through indirect 13C detection, Physical Chemistry Chemical Physics, 15 (2013) 7613-7620.

[32] J. Gregory, Vera circuli et hyperbolae quadratura, 1667.

[33] B. Taylor, Methodus incrementorum directa & inversa, 1715.

[34] S.M. Goldfeld, R.E. Quandt, H.F. Trotter, Maximization by quadratic hill-climbing, Econometrica: Journal of the Econometric Society, (1966) 541-551.

[35] J. Greenstadt, On the relative efficiencies of gradient methods, Mathematics of Computation, 21 (1967) 360-367.

[36] A. Banerjee, F. Grein, Convergence behavior of some multiconfiguration methods, International Journal of Quantum Chemistry, 10 (1976) 123-134.

[37] M. Hebden, An algorithm for minimization using exact second derivatives, UKAEA Theoretical Physics Division Harwell, 1973.

[38] D. Goldfarb, Curvilinear path steplength algorithms for minimization which use directions of negative curvature, Mathematical Programming, 18 (1980) 31-40.

[39] C.J. Cerjan, W.H. Miller, On finding transition states, The Journal of Chemical Physics, 75 (1981) 2800-2806.

[40] J.J. Moré, D.C. Sorensen, Newton's method, in, Argonne National Lab., IL (USA), 1982.

[41] R. Shepard, I. Shavitt, J. Simons, Comparison of the convergence characteristics of some iterative wave function optimization methods, The Journal of Chemical Physics, 76 (1982) 543-557.

[42] J.J. Moré, D.C. Sorensen, Computing a trust region step, SIAM Journal on Scientific and Statistical Computing, 4 (1983) 553-572.

[43] A. Banerjee, N. Adams, J. Simons, R. Shepard, Search for stationary points on surfaces, The Journal of Physical Chemistry, 89 (1985) 52-57.

[44] J. Baker, An algorithm for the location of transition states, Journal of Computational Chemistry, 7 (1986) 385-395.

[45] G.H. Golub, C.F. Van Loan, Matrix Computations, 4th ed., The John Hopkins University Press, 2013.

[46] A. Goldstein, J. Price, An effective algorithm for minimization, Numerische Mathematik, 10 (1967) 184-189.

[47] K. Levenberg, A method for the solution of certain non–linear problems in least squares, (1944).

[48] D.W. Marquardt, An algorithm for least-squares estimation of nonlinear parameters, Journal of the Society for Industrial & Applied Mathematics, 11 (1963) 431-441.

[49] P.E. Gill, W. Murray, Newton-type methods for unconstrained and linearly constrained optimization, Mathematical Programming, 7 (1974) 311-350.

[50] P.E. Gill, W. Murray, M.H. Wright, Practical optimization, (1981).

[51] H. Padé, Sur la représentation approchée d'une fonction par des fractions rationnelles, Gauthier-Villars et fils, 1892.

[52] P. Wolfe, Convergence conditions for ascent methods, SIAM Review, 11 (1969) 226-235.

[53] P. Wolfe, Convergence conditions for ascent methods. II: Some corrections, SIAM Review, 13 (1971) 185-188.

[54] I. Kuprov, Spin system trajectory analysis under optimal control pulses, Journal of Magnetic Resonance, 233 (2013) 107-112.

[55] G. von Winckel, A. Borzi, QUCON: A fast Krylov–Newton code for dipole quantum control problems, Computer Physics Communications, 181 (2010) 2158-2163.

[56] G. Ciaramella, A. Borzì, SKRYN: A fast semismooth-Krylov–Newton method for controlling Ising spin systems, Computer Physics Communications, 190 (2015) 213-223.

[54] See the supplementary information at [URL will be inserted by AIP] for the *Spinach* console logs giving complete spin system and algorithm setting details for the simulations described in the paper.